\documentclass{PoS}

\title{May GWs signals by BH-BH merging be associated with any \protect{$\gamma$} or \protect{$\nu$} burst?
The case of a $NS-NS$ merging in GW-GRB170817A.}

\ShortTitle{May GW be associated with GRB? The GW170817A event}

\author{\speaker{Daniele Fargion}\\
       Physics Department \& INFN Rome1, Rome University 1, P.le A. Moro 2, 00185, Rome, Italy\\
       MIFP, Via Appia Nuova 31, 00040 Marino (Rome), Italy\\
       E-mail: \email{daniele.fargion@roma1.infn.it}}

\author{{Pier Giorgio} {De Sanctis Lucentini}\\
Physics Department, Gubkin Russian State University (National Research
University)\\
65 Leninsky Prospekt, Moscow, 119991, Russian Federation\\
E-mail: \email{desanctislucentini.pg@gmail.com}}

\author{Pietro Oliva\\
      Niccol\`o Cusano University, Via Don Carlo Gnocchi 3, 00166 Rome, Italy\\
      MIFP, Via Appia Nuova 31, 00040 Marino (Rome), Italy\\
      Department of Sciences, University Roma Tre, Via Vasca Navale 84, 00146 Rome, Italy\\
        E-mail: \email{pietro.oliva@unicusano.it}}

        \author{Maxim Yu. Khlopov\\
      Center for Cosmoparticle Physics Cosmion; National Research Nuclear University MEPHI,
Kashirskoe Sh., 31, Moscow 115409, Russia\\
        E-mail: \email{khlopov@apc.univ-paris7.fr}}

\FullConference{MULTIF2017 \\
	 June, 2017\\
	 Mondello, Italy}

\usepackage{journals, amsmath}

\abstract{The Gravitational Wave (GW) events GW150914, GW151226, GW170104 detected by LIGO were a record of  Black Hole binary merging system (BH-BH) very probably  in nearly empty or a vacuum space; such a kind of events will be mostly with no  baryon mass (plasma or dense masses) and therefore mute or blind in any correlated gamma band. By best GW triangulation (as soon as Virgo has been active) their position will be widely located only in a smeared sky (tens or hundred square degree) because of the absence of any correlated spherically-symmetric electromagnetic signal whose photons might be pointing to the exact sources in the sky. If the  GWs events might be born inside a globular cluster, a star forming region or along a spiral AGN accretion disk their additional accreting mass may be the needed baryon load to explode and shine: in those dense places BH-BH collapse  may also offer an optical-X-$\gamma$ afterglow via their baryon lightening and photon tracks. However these peculiar orbiting or multi-body  systems should also imprint  their presence in the inner Kepler period as well as in a perturbed GW signature by unusual time structure. Moreover any   Black Hole active (by a relic jets and-or  an accretion disk) might shine and blaze during the collapse with a BH  by its jet too: however these beams, being extremely collimated, are rarely   pointing toward us during the same brief GWs emission. Only very nearby (tens Mpc) BH-Neutron Star (NS) or $NS-NS$  cannibal merging might be associated with a desired, visible and correlated spherical NS explosion (kilonova one); these rare explosion and their GWs might therefore be localized by photon tracks. But they are lower mass system and they require much lower threshold or just nearer distances.  Because of such nearer cosmic volumes (tens Mpc) and because of the very anisotropic beamed GRB associated, these kilohertz event are possibly still very rare and unexpected in gamma sky (even with  LIGO Virgo array sky narrow view).
 However the very exceptional GW170817 GRB170817a event last August 2017 it took place and it was related to a very first $NS-NS$ collapse
in a SN-kilonova spherical explosion. This event had been recorded both by its few (day and months) optical transient but as well as by its weak, prompt, short, GRB, two seconds later the same LIGO-Virgo detection. The GRB170817a gamma, X, radio signature  it was exceptional  in many features and one may wonder and he may ask if it was really correlated to the GWs. Its unique values (softer, weaker, the most near one ) made GRB170817a very possibly an off-axis jet detection. However there are solid arguments that suggest that such a GRB are not just blazing within a collimated beam jet but that they are also shining in a wider spread gamma equatorial disk blazing, orthogonal to the jet itself. In a few words we were observing the event not along its jet but mostly orthogonal to it. This GRB170817a geometry may better explain the otherwise unexpected beamed to us event. In this paper therefore we summarize the astrophysical and the cosmological signature of such a long desired   multiple astronomy.
}

\begin{document}

\section{Introduction} \label{sec:intro}
This article consider the main contribute of recent years LIGO-VIRGO GWs detections.
We address toward two main questions: How sharp GW astronomy might be observed and how much relevant are the LIGO
GW backgrounds in a known cosmological density frame?
Indeed in  the following \\
$1)$first section we discussed as a first approximation  the cosmological energy and number fluxes
based on the few BH-BH and NS-NS GWs observation in  the last few years.\\
$2)$In the second section we focalized to the GW and its neutrino traces as well as to the expected gamma signal in NS-NS or NS-BH collapse.\\
$3)$In the third section we remind the additional indirect signal of GWs made
by GWs conversion (via galactic or planetary magnetic fields)
into a tiny kilo-Hertz radio bangs.\\
$4)$In the fourth section we considered the BH possible origin.\\
$5)$In the fifth section we discussed the main nature of multi-currier signal in NS-NS versus GRB.\\
In particular we briefly analyzed the recent GRB 170817a discover and its probable new peculiar jet morphology.\\
$6)$In the final conclusion section we summarized the GW future opportunities in view of a revival of LIGO-VIRGO detection
and in consideration of a wider visibility of GRB explosion and its  jet spread persistence both in a gamma, X, optical and radio disk orthogonal to the thin more collimated, precessing GRB jet.

\subsection{A BH-BH collapse making a GWs Astronomy? The present  cosmic GW legacy}

The  advanced Laser Interferometer Gravitational-Wave Observatory (LIGO) \cite{Aasi et al.(2015)}
 identified two binary black hole coalescence signals two years ago
with high statistical significance $\sigma$ or Signal$/$Noise,  in a summary below:
\begin{equation}
GW151226\rightarrow    5.3   \sigma \cite{Abbott et al.(2016b)}
\end{equation} 
\begin{equation}
LVT151012\rightarrow  S/N 9.7;\rightarrow  1.7 \sigma 
\end{equation}
\begin{equation}
GW150914\rightarrow    5.1 \sigma \cite{2016PhRvL.116f1102A} 
\end{equation}
\begin{equation}
GW170104\rightarrow   S/N 13  \cite{Abbott et al.(2016c)} 
\end{equation}
\begin{equation}
GW170814\rightarrow   S/N 18   \cite{Abbott.et.al.(2017)} 
\end{equation}
\begin{equation}
GW170817a\rightarrow  S/N 32.4 \cite{Abbott.et.al.(2017c)} 
\end{equation}

 The last GW170104 source is a heavy binary black hole system,
with a total mass of of 50 solar ones located at far ($z\simeq 0.2$) cosmic distances (farthest confirmed event to date).
More recent BH-BH merging as in GW170608, and in particular in GW170814 observed also by Virgo array,
strongly confirmed the ability to record the GWs, but with no electromagnetic signal.
Only the most recent an d spectacular GW170817, discussed in last section, had the first
multi-currier signature both in GW and in electromagnetic field.

Astronomy is, etymologically, the art of identifying sources in the sky and to collocate them with their names toward the (possible) optical  or electromagnetic (EM) counterpart. Many potential astronomy today are still blind or invisible or at best myope: the relic SN neutrinos (yet to be discovered), the few tens TeV-PeV IceCube neutrinos (not yet correlated to any source), the same Cosmic Rays (CR) bent by galactic magnetic fields or even the Ultra High Energy (UHECR) whose rigidity leads to a myope and-or a coarse astronomy possibly born in recent years \cite{Fargion:2018}, \cite{Fargion:2015}:
(Very recent clustering seem to point to a Local Universe ruled by few star-burst AGN sources as Cen A, M82, NGC 253)

The recent first GW detection by LIGO \cite{2016PhRvL.116f1102A} on September 14\textsuperscript{th} 2015,
 it does offer the first view of the huge BH-BH GW energy density or its energy fluency, but not yet any clear source localization.
The nominal first event energy fluency and its power fluency (assuming at 400 Mpc distance and assuming a conservative ``one a year'' event rate, well calibrated with more recent \cite{Abbott et al.(2016c)}) leads to
\begin{equation}\label{eq:nom}
\frac{\Delta E}{4\pi R^2_{400\,\mathrm{Mpc}}}\simeq\frac{2\cdot3\cdot10^{33}\cdot9\cdot10^{20}\,\mathrm{erg}}{4\pi\,
\left(1.2\cdot10^{27}\,\mathrm{cm}\right)^2}\simeq0.3\,\mathrm{erg}\,\mathrm{cm}^{-2}\,\mathrm{sr}^{-1}
\end{equation}

\begin{equation}\label{eq:Eflux}
\frac{\mathrm{d}E}{\mathrm{d}A\, \mathrm{d}\Omega \, \mathrm{d}t}=\frac{0.3\,\mathrm{erg}}{3\cdot10^7\,\mathrm{cm}^{2}\,\mathrm{s}\,\mathrm{sr}}\simeq 1.25\cdot10^{4}\;\mathrm{eV}\;
\mathrm{cm}^{-2}\,\mathrm{s}^{-1}\,\mathrm{sr}^{-1}.
\end{equation}

\begin{figure}[t]
\begin{center}
\includegraphics[width=0.99\textwidth]{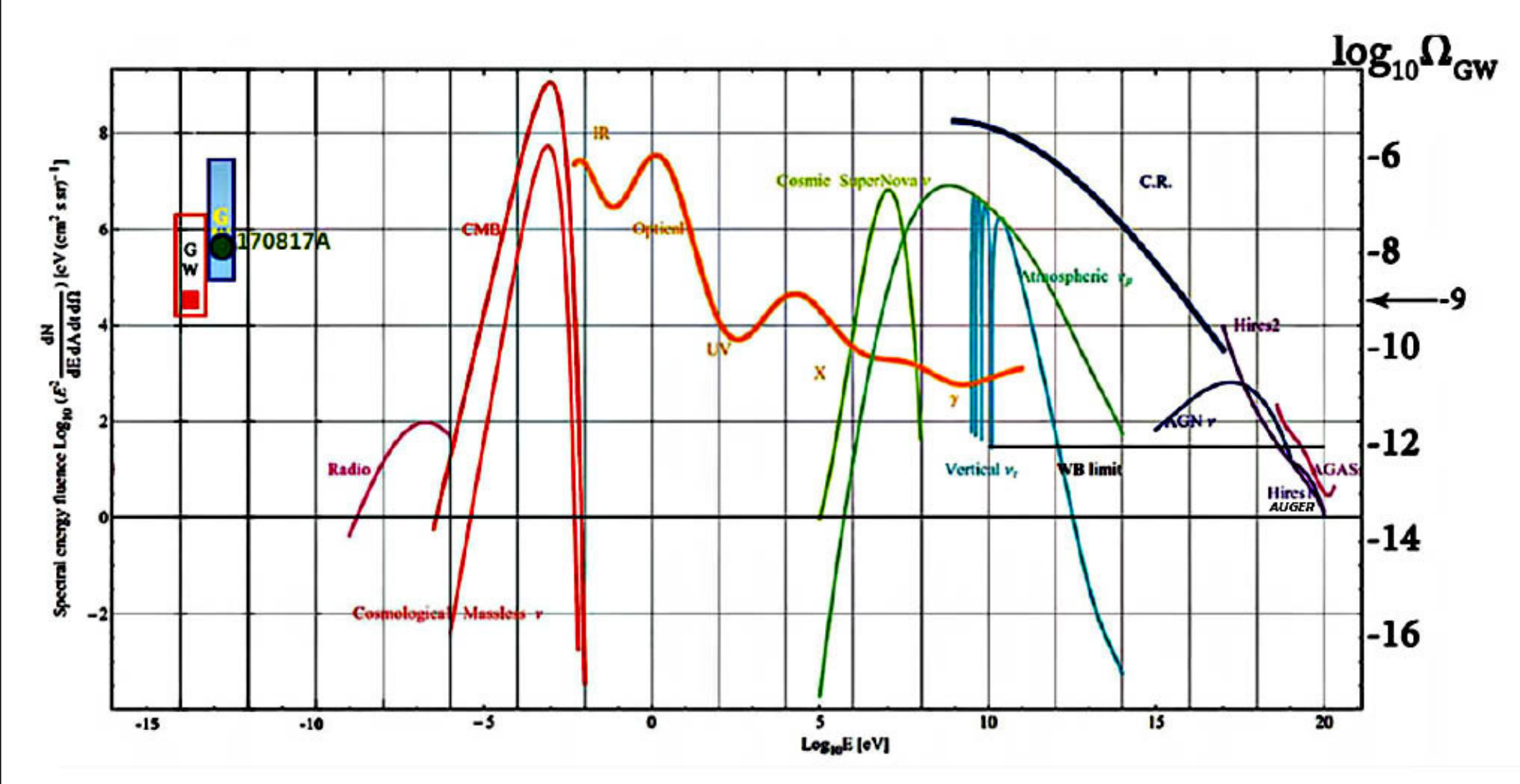}
\caption{The cosmic energy fluency spectra in logarithmic scale as a function of the particle energy also in logarithmic scale.
The huge solar contribute and the high optical galactic plane noise is here ignored. The Big Bang infrared signal is the ruling
energy contribute corresponding to nearly $4.7\cdot10^{-5}\,\Omega_{c}$. The observed LIGO GW in red area  and the very probable associated kilohertz additional kHz noise (cyan area) has been marked in approximated box windows. As in the text these energy fluency overcome the corresponding EM ones in gamma and radio by two or more order of magnitude. Note that the optical and infrared e.m. contribute are mostly born in local Universe. Note the relic  SN neutrino signals at tens MeV energy whose detection is under rush search in SK (gadolinium implemented) detector, it is somehow comparable to Big Bang infrared signal.}
\label{fig:fig1}
\end{center}
\end{figure}

It might be questioned if such a huge cosmic energy density fluency of $\sim0.3\,\mathrm{erg}\,\mathrm{cm}^{-2}$ and its apparent brightest power in 0.4~s duration burst over-passed any previous observed radiative one. The unique corresponding energy density fluency has been observed\cite{2005PASJ...57L..11Y, 2005Natur.434.1104G} during the giant flare from SGR 1806-20, on December 27\textsuperscript{th} 2004 (just to remind: this was just a day before the largest tragic tsunami in Sumatra, Malaysia). The gamma satellite RHESSI and its particle detector data implied\cite{2005Natur.434.1098H} a spike of fluence in photons  at $\geq30$~keV energy as large as ($1.36\pm0.35$)~erg~cm\textsuperscript{-2}. However this SGR event was a galactic one, not a cosmic one. Moreover, just for an historical comparison, the nearest SN 1987A and its neutrino burst from Large Magellanic Cloud (LMC) has been even much more energetic\cite{1987PhRvL..58.1490H} than the SGR 1806-20 and LIGO GW150914, by nearly a million times. However as being a cosmic event LIGO GW150914 overcame any other GRB or AGN cosmic flare event detected in last century.
Also the consequent probable averaged  GW  flux number and its energy fluency will be enormous respect the EM ones.
The LIGO collaboration has evaluated its first  GW signal as indebted to a wide range of possible cosmic rate:
 2--400~Gpc\textsuperscript{-3}~yr\textsuperscript{-1} where the minimum value of 2~Gpc\textsuperscript{-3}~yr\textsuperscript{-1} corresponds to a much (a dozen time) rarer than previous one rate (see equation (\ref{eq:Eflux})) and at an averaged lower power energy fluency of
\begin{equation}
\frac{\mathrm{d}E}{\mathrm{d}A\, \mathrm{d}\Omega \, \mathrm{d}t}\simeq2\cdot10^3\;\mathrm{eV}\,\mathrm{cm}^{-2}\,\mathrm{s}^{-1}\,\mathrm{sr}^{-1}
\end{equation}
which is to be considered as the most ``conservative prudential rate''. The two additional events may in average confirm this typical distance. The low GW energy for each characteristic graviton has a frequency on average of $\langle f\rangle_{\mathrm{GW}}=150$ Hz (and a peak at 250 Hz) leading to a very low energy $\mathrm{E}_{\mathrm{GW}}\simeq6.2\cdot10^{-13}\,\mathrm{eV}\left(\frac{\langle f\rangle_{\mathrm{GW}}}{150\,\mathrm{Hz}}\right)$ implying an abundant flow of gravitons during the one year rate or 2 Gpc\textsuperscript{3}~yr\textsuperscript{-1}:
\begin{equation}
\frac{\mathrm{d}N}{\mathrm{d}A\mathrm{d}\Omega \mathrm{d}t}\simeq3.2\cdot10^{15}\left(\frac{\phi_{\mathrm{GW}}}{2\,\mathrm{Gpc}^3\,\mathrm{yr}^{-1}}\right)\left(\frac{\langle f\rangle_{\mathrm{GW}}}{150\,\mathrm{Hz}}\right)^{-1}\mathrm{eV}\,\mathrm{cm}^{-2}\,\mathrm{s}^{-1}\,\mathrm{sr}^{-1}
\end{equation}
This estimated graviton number flux exceed the  2.7~K cosmic thermal photon number flux by a hundreds times making GWs the most abundant massless cosmic messenger in our universe. Actually,  GW in the afterglow of the Big Bang or the very same ``virtual'' graviton that is keeping the reader sitting on his chair is by many more orders of magnitude the most abundant massless messenger; for an analogy just remind also the well known terrestrial magnetic fields whose virtual photons (huge dense number) are bending our compasses and the Cosmic Rays flight in the Universe; see figure~\ref{fig:fig1}.

\begin{figure}[t]
\begin{center}
\includegraphics[width=0.98\textwidth]{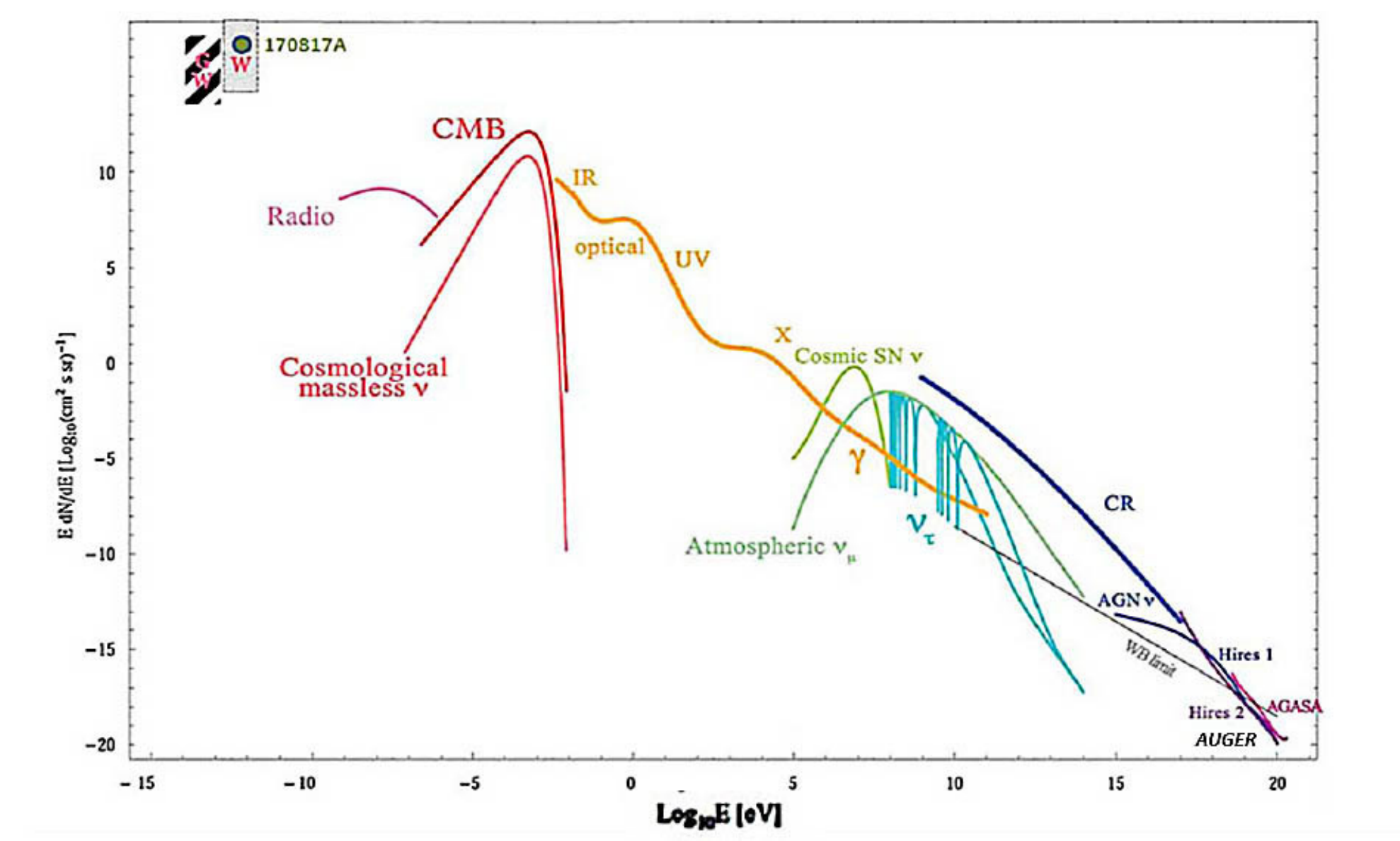}
\caption{The cosmic number spectra in logarithmic scale as a function of the particle energy also in logarithmic scale.
The huge solar contribute and the high optical galactic plane noise is ignored. The Big Bang infrared signal is ruling. The observed LIGO GW is the dashed area  and the very probable associated kHz additional noise (gray area) has been marked in a box windows. As in the text these number's contribute overcome the corresponding EM ones in gamma and radio. Note that the optical and infrared contribute are mostly born in local Universe. It will be surprising that such huge GW number flux won't be visible in astronomy but without a correlated EM afterglow  information over direction will be almost lost.}
\label{fig:fig2}
\end{center}
\end{figure}

The more probable binary (a smaller size BH systems, whose smaller mass is of the order of $\sim3\,\mathrm{M}_\odot$ ) may be merging at higher frequency at a peak of $f=2.5$~kHz: these events may contain NS companion whose masses collapse and explosive death are detectable; they are detectable in a  narrow located  distances by an order of magnitude respect to GW150914 sources and their threshold detection volume is consequently a thousand time smaller; their larger  abundance  (by a reasonable spectra mass dependence)  might compensate (or not) their nearer detection volume leading (or not) to their discover in a near future. Their optical correlation might then be possible. Nevertheless the 30~M$_\odot$ BH binary systems and their GWs seem to be a quite dominant signal in the overall cosmic energy fluency at LIGO three events as shown in figure~\ref{fig:fig2}.
Indeed, the unique  detection by LIGO of a short $NS$ $NS$ collapse, discussed in next sections, implies a definite narrow windows of GWs of hundreds and thousands Hz, whose fluency are anyway already remarkable in comparison with all the other EM radiations,(see the last section).

As a matter of fact, only the cosmic Black-Body Radiation (BBR) at 2.7~K is still a dominant one ($\phi_{\mathrm{BBR}}\simeq10^8$~eV~cm$^{-2}$~s$^{-1}$~sr$^{-1}$) as well as the probable and  ``soon to be detected'' $\phi_{\mathrm{SN}}$ relic from all integral Super Nov{\ae}, the Cosmic SN $\nu$ in figure~\ref{fig:fig1} and in figure~\ref{fig:fig2}, which is  about $\phi_{\mathrm{SN}}\simeq10^7$ eV cm$^{-2}$ s$^{-1}$ sr$^{-1}$. For instance, the most celebrated Gamma Ray Burst (GRB) fluency, in average by Compton satellite, is  $\phi_{\mathrm{GRB}}\simeq10$~eV~cm$^{-2}$~s$^{-1}$~sr$^{-1}$, similar to the UHECR one $\phi_{\mathrm{UHECR}}\simeq10$~eV~cm$^{-2}$~s$^{-1}$~sr$^{-1}\left(\frac{\mathrm{E}_{\mathrm{UHECR}}}{10^{19}\,\mathrm{eV}}\right)^{-2}$, it is usually referred as  the Waxmann Bachall (WB) bound it is hundreds or thousands times below the LIGO GW one. The natural consequence is that if a tiny (few percent) component
of the GW radiation is transferred to EM waves than the correlation GW-EW signal might be found easily. The GRB with a powerful transient luminosity comparable to $0.1$\%  are the best candidate transient event to be associated. Does this connection occur on GW170817-GRB170817a?

\section{A LIGO GW  connection to GRB?}

From above there are many reasons to foresee a fast e.m. transient associated to these LIGO-VIRGO astronomy. Indeed there have been a huge attention and attempts to find correlated gamma signals with GWs, with no success.
However as we have mentioned the BH BH merging in vacuum is a source only (or mostly) of \emph{silent EM} GWs with no EM tail.
Naturally binary collapse in  AGN accretion disk or in dense globular cluster may shine in e.m. waves, but their
frequency might be rare and their GW signature will be somehow spoiled and disturbed by the third body system and-or mass transfer. Therefore
these dense surrounding might be traced in the  GW signal anyway.
Moreover the surrounding baryon screen will transfer into a gamma signal within an opaque dense wall leading to a  long (hours) delay: this is  due to baryon photon opacity and its over-Eddington dense electron pair screening.
For this reason the very timed (and controversial) gamma flare correlation with  LIGO GW150914, event found in FERMI with $0.4$~s. detector delay, it seems very improbable.
However just during the very first $NS-NS$ GW event on 17 August 2017 there have been an outstanding and exceptional GRB 170817A, whose discussion require a additional sections just below and  at the  article end.

\subsection{$NS-NS$ or $NS-BH$ collapse GRB?}

A more noisy neutron binary system while in collapse might be source of GW and also of the associated SN like explosion.
However the masses are smaller, the consequent GW signal is much weaker and at higher frequency,  the tidal disruption may smear  and dilute the  GW emission.
Also a BH-NS fast cannibal collapse may lead to a very eccentric trajectory encounter or in the late stage of the NS strep-tease to sudden NS instability, causing GWs and an associated NS like a SN correlated explosion \cite{Fargion:2016}.
These events, because a ten times smaller masses, might be observable at much lower (about ten times) intensity and  at a consequent much nearer (ten times) distances and at a higher (kHz) frequencies. This foreseen occurrence  had been met on GW170817 event, discussed next section.  They ($NS-NS$, $NS-BH$ merging) are the  best candidate for a visible, mostly in optical way, to a correlated GW-SN like  explosion  \cite{Fargion:2016}. The possible UHECR trace are not much correlated observable signals (in our lifetime) because of the slow random walk of charged nuclei or even proton, that will held for century to come in a very delayed time scale.

\subsection{A GW spherical signal versus a jetted GRB one}

Somehow  any discover of any (we believed improbable) LIGO GW  correlated GRB event  it will be a winning signature of the old spherical explosive GRB Fireball model.
Indeed the early GRB model has been a very spherical huge explosion: the celebrated Fireball model \cite{Cavallo01071978}, \cite{1986ApJ...308L..47G},
 \cite{1986ApJ...308L..43P}, \cite{1992ApJ...395L..83N}, \cite{1994ApJ...430L..93R}, \cite{1993ApJ...405..278M} on last decades 1980--2000 years. In our-days this spherical model is no longer alive. It has been de-throned by fountain Fireball, whose beam range ten or few degree.  However if a spherical GRB would occur, with its associated GW merging, than the EW observation must occur immediately because both GW explosion as well as the GRB burst are almost spherical symmetric. Only a time delay may separate their connection but such connection will be easily found soon. However the GRB has different model today.
Indeed, the very high GRB time variability, the huge isotropic power needed and the consequent narrow size of the source
 (and its high over-Eddington opacity for such GRB hard brightrst gamma spectra events) forced most the authors (since 2000s) to consider more and more a beamed fireball-fountain \cite{1997ApJ...485L...5W, 1997MNRAS.288L..51W, 2016ApJ...818...18G} models whose opening angle is at least as wide as $\frac{\Delta\Omega}{\Omega}\simeq10^{-3}$. On the other side we offered since the earliest GRB-SN event on April 1998  an even more beamed, persistent model: a very thin jet, blazing and persistent  model\cite{1999A&AS..138..507F, 2001foap.conf..347F, Fargion:2001xf}, \cite{Fargion:2016} $\frac{\Delta \Omega}{\Omega}\simeq 10^{-6}-10^{-8}$. This jet is spinning and precessing with a decaying output, able to fit by its blazing geometry the early erratic  GRB luminosity as well as to explain the  very puzzling variable re-brightening  of GRB  afterglow or their rare, otherwise unexplained, X-ray precursor\cite{Fargion:2006we}. In particular most recent version \cite{Fargion:2016} based on a NS strep tease by a BH merging guarantees an early electron pair jets (with no neutrino correlated signal) and a late NS fragmentation leading to SN like explosive event. These model explain  the rare but puzzling and critical GRB and SN coexistence.

 Anyway beamed GRB (Fireball or precessing Jet) are not spherical at all. Therefore the discover, soon or later, of LIGO-VIRGO GW-GRB connection is a benchmark of the GRB model: a spherical GRB will be soon be observed in gamma connection with its GW; a jetted one will be rarely or mostly not observed with the its originating GW. In a more  quantitative estimate, the GRB beam pointing inside a solid angle of $\frac{\Delta \Omega}{\Omega}\simeq 10^{-3}$ will need hundreds of events and tens LIGO-VIRGO recording years (even more for our thinner GRB jet model \cite{Fargion:2016}). The LIGO-VIRGO GW visibility is somehow turned in this Amletic question: GRB, beaming or not beaming jet?
as we mentioned only some nearer GRB fed by NS-BH  or $NS-NS$ merging may also trace by a parasite SN like explosion in a correlated spherical event. Not the BH-BH merging in a vacuum space. As mentioned above, we believe that the jet beam in GRB is quite narrow, below a few micro steradian so that it may spray
in long times in little wider cones.  In our view GRB  is a persistent jet that it may arise as a precursor (of the same main blaze GRB) and it  may occur earlier (days, weeks months) before  than the GW event where one of the BH jet and its companion are merged in unique BH.
 This implies that in general the observation of a GRB in axis occurs at a very rare rate (once every thousands or more) appearing mostly off axis or not at all. The contemporaneous GWs it is therefore very improbable.
 We now consider an additional tool in GWs detection and BH sources; later on we return to the peculiar GW-GRB connection in NS-NS collapse.

\section{The Photon Graviton conversion}

The foreseen kHz to-be-observed-soon GW are reminding us of a relevant connection with a parasite EM radiation: the parametric photon-graviton conversion.
 The G-$\gamma$ conversion, were first considered in early 60s\cite{Gertsenshtein1961} and studied through the last decades \cite{Mitskevich}, \cite{Boccaletti1970}, \cite{Dubrovich(1972)}, \cite{1974JETP}, \cite{Fargion1995}, \cite{Dolgov.et.al.(2017)}. Gertsenshtein first argued that the coupled gravity-EM energy equation will generate EM waves.
 Historically, gravity bending light was first proposed by Einstein himself, exactly a century ago. The same  Feynman diagram (a real photon plus a virtual graviton of the gravitational object are leading into a scattered photon again), may occur in the opposite way (a real graviton plus a virtual photon are leading to a real photon).
 The conversion process occurs because, while GWs  propagate and squeeze in space-time, they also force the intergalactic magnetic field lines to tremble and vibrate, thus producing a secondary tuned EM waves.

The G-$\gamma$ conversion process can be interpreted as a secondary effect of gravitational synchrotron radiation\cite{Ge62}
and it has been also analyzed by different authors, \cite{Lupanov.1967}, \cite{Sushkov.Khriplovich.1974}, \cite{DiambriniPalazzi1987}, \cite{Fargion.1991}, \cite{PhysRevD.87.104007}. Let us briefly recall the equations of the GW $\rightleftarrows$ EW oscillations according to Landau and Lifshits's notation\cite{LANDAU1975}: let's consider a nearly-flat space-time $\eta_{ik}$ with a small metric perturbation
$h_{ik}$ and  let's call the traceless tensor $\psi^i_k$. We can then write
\begin{equation}
g_{\mu\nu}=\eta_{\mu\nu}+ h_{\mu \nu};  \;\;\;\;\
h_{\mu\nu}=\psi_{\mu\nu}-\frac{1}{2}\psi^{\sigma}_{\sigma}g^0_{\mu\nu},
\end{equation}
so that the Einstein field equations $ G_{\mu \nu}=(8 \pi G/c^4)T_{\mu \nu}$
in the linear approximation become
\begin{equation}\label{LL1}
\square\,\psi^{\mu}_{\nu}= \square\,h^{\mu}_{\nu}
=-\frac{16\pi G}{c^4} \tau^{\mu}_{\nu}        
\end{equation}
where $\tau^{\mu}_{\nu}$ is the energy momentum tensor (the first equality
in equation (\ref{LL1}) holds because $\psi^{\mu}_{\mu}=0$). In an external stationary
EM field $F^{\sigma \tau(o)}$ and in the presence of a free EW,
$\tilde{F}^{\sigma \tau}$, the total EM field is
\begin{equation}\label{LL2}
F^{\mu \nu}\equiv F^{\mu \nu(0)}+ \tilde{F}^{\mu \nu}   
\end{equation}
The corresponding energy-momentum tensor is then
\begin{equation}
\tau^{\mu}_{\nu}=\frac{1}{4\pi} \left[F^{\mu \sigma} F_{\nu \sigma}-\frac{1}{4}
   \delta^{\mu}_{\nu}\left(F^{\sigma\tau}F_{\sigma\tau}\right)\right]
\end{equation}
and
 \begin{equation}
\square\,h^{\mu}_{\nu}=-\frac{16 \pi G}{c^4} \tau^{\mu}_{\nu}=-\frac{8G}{c^4}
\left[F^{(0)\mu \sigma} \tilde{F}_{\nu \sigma}- \frac{1}{4} \delta^{\mu}_{\nu}
\left(F^{(0)\sigma \tau} \tilde{F}_{\sigma \tau} \right)\right].  
\end{equation}

 In a few words the spacetime squeeze the magnetic stationary fields making them to emit
and the photon with the stationary magnetic field couple and makes energy trembling and generating GW:
 This idea was rediscovered while we  proposed (by one of us)\cite{DiambriniPalazzi1987} a twin artificial experiment , firstly producing ``optical'' waves in magnetic tunnels then high energy graviton  and later on getting back observable photons;  unfortunately this double EW-GW-EW twin conversion is very poor, today, and it is not feasible. This parametric conversion in low radio band in astrophysics  is somehow modulated and diluted by the plasma presence that, in a way, makes photons behave as they had  a mass. Therefore this GW-EW conversion in space has the damping related to the free charge in space.
   The conversion $\langle\alpha\rangle$ do not grows quadratically $\langle\alpha\rangle\simeq\frac{G}{c^{4}}\cdot B^{2}L^2$ but just linearly:
   \begin{equation}
\langle\alpha\rangle\simeq\frac{G}{c^{4}}\cdot B^{2}L_{\mathrm{Coherence}}L_{\mathrm{Distance}}
\end{equation}
   A first conclusion follows: there is much more GW radiation in our universe than the high energy astrophysical one, but these huge low frequency kHz signals has a trace anyway in a tiny radio bang waves \cite{Fargion1995} at milli-Jansky threshold fluency. We wonder if these weak  signals may be observable in future experiment.
   We would like to notice that underwater submarine may communicate in these lowest radio frequency. Therefore it might be of interest to verify if any military antenna at ground or better in space may have been recording such a radio bang in correlated GWs event.

\section{Black Hole progenitors}

There is a negligible probability that two single large $\approx 30\,\mathrm{M}_{\bigodot}$ will interact and bind and merge
as the GW150914 merging event in an average dense cosmology, even in the extreme unrealistic assumption that these BH are the total missing dark matter of the Universe: in the whole history of the Universe this may occur only once (for reasonable speed velocity) much below unity ($\simeq 10^{-4}$) or just few times in a dozen billion years if one assume that these dark matter BHs are clustered  in million times denser galaxy volumes.
There are other places that may be a catalyzer for such rarest encounter: the inner dense medium of largest giant star (three body event) or along accretion disk or halo of giant AGN BH \cite{Fargion:2016b}. However these possibilities sound much more rare of a more probable genetic parental birth: the largest giant star binary systems. Such a hottest and most massive double star has been recently discovered in the LMC.

\subsection{The Giant star system VFTS 352 origin of BH}

The double giant star system VFTS 352 is located about $160\,000$ light-years away in the Tarantula Nebula in LMC. This remarkable region is the most active nursery of new stars in the nearby Universe and new observations from ESO and VLT  have revealed that this pair of young stars is among the most extreme and strangest yet found. This system
 VFTS 352 is composed of two very hot, bright and massive stars that orbit each other in little more than a day. The center  of the stars are separated by a dozen of million of kilometers. Their surfaces overlap and a bridge has formed between them.  VFTS 352 is not only the most massive known in this tiny class of over-contact binaries it has a combined mass of about $57$ times that of the Sun, comparable with the GW150914 binary BH masses. The rarity of such huge giant stellar system stand in front on more mundane and abundant (and less giant) large stars system able anyway to lead to few (3-6) solar mass BH. Therefore there must be a much larger population of  lighter (3-6) solar mass BH binary systems whose detection metric deflection is nearly ten times weaker and whose  consequent detection distances are ten times smaller inside a  nearby Universe.

\subsection{Future few solar masses BH GW event at kHz}

Therefore the LIGO  result may soon lead to detection of nearer  $3\,\mathrm{M}_\odot$ binary systems masses as it may be better observable via threefold LIGO-Virgo array detection already during  few next years or decade. These events would occur in a narrow universe ($\lesssim40$ Mpc), possibly within the same GZK cut-off as it is for the UHECRs. Naturally, while one might barely hope for an EM afterglow detection, as it has been widely tried and searched for identification (we actually believe, as shown below, that it will be highly improbable), there won't be any possible correlation with the quite lazy and slow and bent UHECR arrival signal hundreds or thousand years later. These expected EM events  of few solar masses BH could then be better observed by future higher sensitive LIGO-Virgo array, inside a much narrower universe possibly also correlated to richest dense galaxy clusters. The possibility to reveal a nearby galaxy cluster correlation (like pointing to Virgo or Coma, but not to the individual galaxy or star source) for few solar mass BH system, seem to us the maximal hope of present array GW detectors.

\section{A LIGO-VIRGO GW  connection to GRB?}

If one consider the huge power of the LIGO BH event one may remind the comparable GRB (apparent) huge
bursting luminosity. Therefore it is natural to attempt  a LIGO  connection to EM or neutrino signal
somehow correlated to GRB or AGN like flare. At the present no TeV neutrino signal has been found correlated with a GRBs.
If the  GRB is a blazing ultra-relativistic jet than the inner cone are the most energetic and most collimated
component of the jet. The UHE neutrino will be mostly in a thinner hard core of the jet. Therefore the first neutrino event may
be probably a precursor of a later  secondary (by wider cone gamma jet) gamma signal.
If the Gamma correlation will be absent any harder (and more collimated) TeV signal sound even much more improbable. Of course
the possibility for such connection (GW-GRB) is related to the binary model and to the expected GRB ejection model.
In the far past (let's say until 1999) most models suggested, that GRB were just an ideal spherical symmetric fireball:
the nearly spherical symmetric GW signal would or could correlate, soon or later, with the spherical Fireball
at same power as a natural candidate source.
However the Fireball spherical model is fade away  for several reasons  and most model imagine GRB
as a beamed jet more (as we suggested $0.1^\circ$) or less ($5^\circ$) beamed and collimated cone.
In this optics the possibility that a GW (spherical) event does correlate with a beamed
GRB jet  (usually pointing elsewhere) seem quite unlikely. In case of the GRBs searching for an afterglow (X,optcal,radio) the opposite was
taking place: a very beamed MeV photon signal was able to hit anyway  wide angle view gamma satellites;
 then  one could easily follow  (after Beppo-Sax,Swift..) the X ray afterglow disentangling its inner optical transient.
  Moreover for GW event there is also a very probable time lag between the its explosions
and any correlated erupting signal, that we underline, it is very possibly, pointing elsewhere.
However, as in next section, such a GW-GRB connection did take place last year.

\subsection{The GW170817-GRB170817A exceptional event}

The very recent discover of a first GW170817  by LIGO-Virgo, an event probably indebt  to a NS-NS collapse, it
correlated with a very rare GRB170817A. Such a connection poses  several questions.
Indeed the same GRB170817A event exhibited several rare and
extreme values all at once:
(a) it is the nearest (among several hundreds) short GRB (SGRB) ever detected (40Mpc)
(b) it is (among several hundreds) the weakest fluency SGRB ever recorded
(c) it is (among several hundreds) the most soft SGRB in whole known SGRB catalogs
(d) it is the first NS–NS (or more rare NS–BH) merging GW event ever detected
(e) it is the nearest GW event observed
(f) it is the first NS–NS merging into a new born NS or a new BH
(g) it is (probably) among most powerful energy fluency GW ever observed
(h) it is the very first GWs correlated with an  afterglows ($\gamma$, X, optical).
(i) it is (among several hundreds) the very first GRB identified not via its $X$ ray afterglow, but via its day after optical transient.
(l) it is (among several hundreds) the most long life GRB brightening (or re-brightening) up now ($\sim 200$ day later).

 Giving the very surprising (rare) statistical  weight of  each of these exceptional parameters, one may reach
 the conclusion that the  GRB time link  and the (Fermi) wide sky solid angle correlation, they  cannot by themselves be compelling at all
 in forcing such a GRB-GW link. Indeed if within  a radius of 40 Mpc (GRB170817a distance) one may contain
 (let say) $10^4$ galaxies, if the NS-NS event occurs once every $10^4$ years, than the NS-NS collapse may take place
 within a second time windows by chance only once every $ 3\cdot 10^{-8}$. If the time windows enlarge to  minutes or hours the correlation factor will be
 be diluted just to just a part over a million. Moreover the simple product of the enlisted statistical weight
 in  values  (a-b-c-i-l), assuming for each $\sim 10^{-2}$ average value, it lead to a suppression factor (probability to occur) $\sim 10^{-10}$.
 This  values as well as the geometrical beaming argument expressed above makes the same correlation quite a weak or even meaningless discover.
 However it is just the same  peculiarity of all above values that it makes  more acceptable the possibility of an off-axis view
 of a GRB \cite{Fargion.Sept.(2017)}. This has been recognized, later on and with all the detailed GW-GRB data, by most authors \cite{LIGO FERMI (2017)}.
 The idea that allow the wider view of a lateral GRB has been named with an ad-hoc term: \emph{the structured jet}. We did found a very different
 road to enlarge the GRB observability, leading to a natural spread signature of the event.

\subsection{The GRB170817A jet and its equatorial gamma disk}
\begin{figure}[t]
\begin{center}
\includegraphics[width=0.99\textwidth]{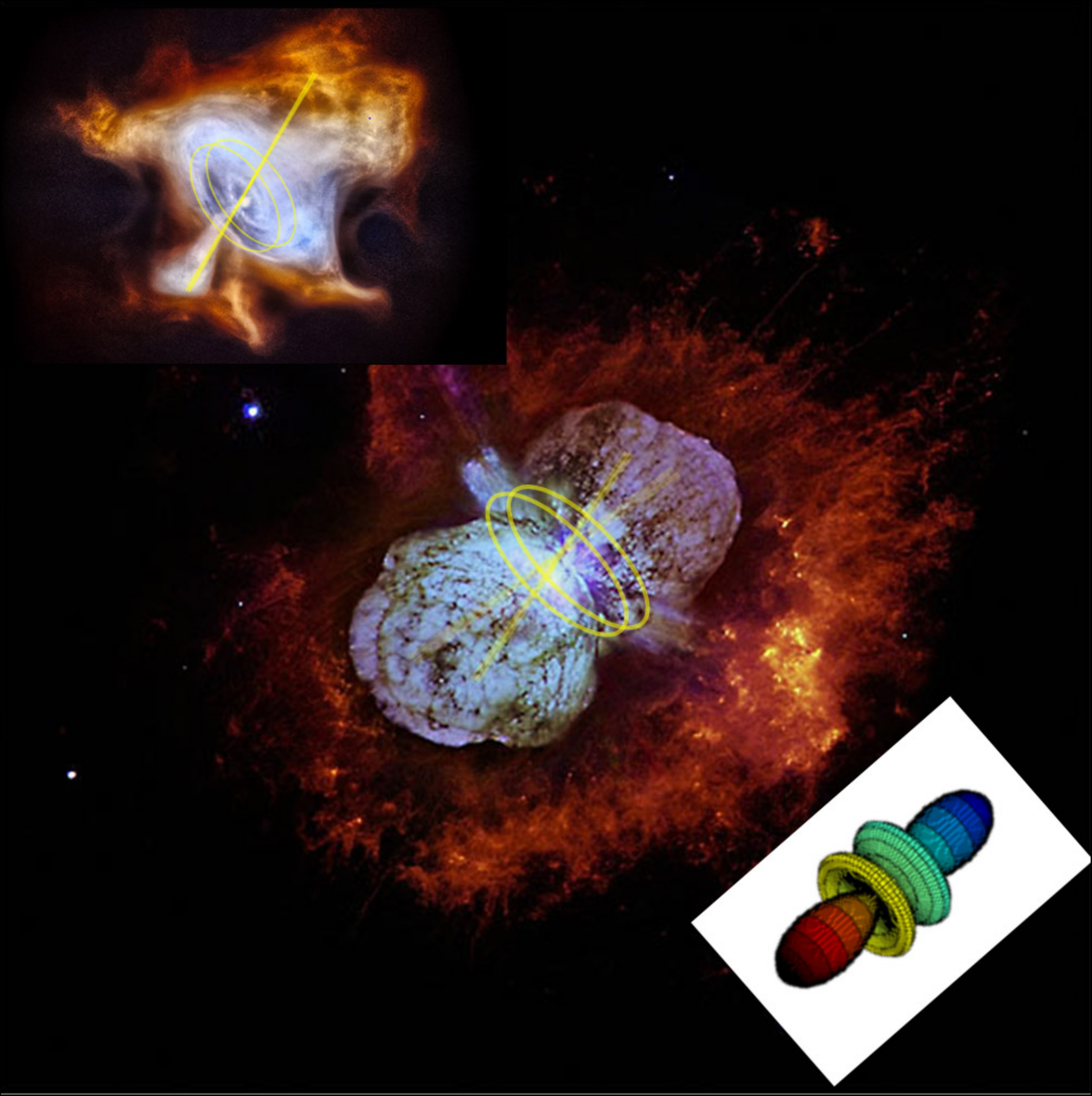}
\caption{Crab Nebula on the top left side: we suggest that the Crab nebula powered by an old gamma pulsar resemble by its geometry  a spinning and precessing jet that
it is also shining (at the equatorial edges by their spirally UHE electron in a jet column)  a twin orthogonal gamma disk. In analogy, at the figure center, we suggest that Eta Carina nebula show similar geometry. Its morphology suggest an old hour glass fed by a twin jet and its orthogonal spread disk it is powered, possibly, by its orthogonal twin gamma disk. We see in these pictures the relic gas ejected and pressured and illuminated by those gamma radiation.
This geometry might resemble  the GRB170817a one; we, the terrestrial observers, have been located along the wider gamma disk solid angle and not along the very narrow thin jet cone as it occurs on the other cases, in most far cosmic, aligned, hard and brightest GRB. On the right side bottom an exemplified multi-pole figure that summarized the gamma luminosity profile and its spinning and precessing signal in a very reduced scale: the inner jet cone might be collimated (even if precessing, spread and and persistent)  many millions times than spherical shells while its thin gamma disk may be shine in a much wider solid angle offering a much higher probability to be observed but at a much less amplified (just from tens to a thousand times) than any spherical explosive luminosity.}
\label{fig:3}
\end{center}
\end{figure}

Indeed for us the idea of an off axis view  (of a thin gamma jet) alone  was (and it is) not enough to encompass the statistical probability to be
at the first GW-GRB connection, within the rare correct tuned  solid angle
 of view (assuming a gamma jet born by electron in Lorentz factor of a hundred or above).
 For this reason we suggested recently a more natural role of the circular spiral jet shining at a plane $\gamma$ disk
 orthogonal to the same jet beam. \cite{Fargion:2017b} (as shown \ref{fig:3}). This geometry amplify the visibility of the GRB170817A by nearly three order of magnitude respect
 to a view along the thin collimated jet beam. The GRB we model are thin spinning, precessing jet; the consequence of such a geometry it is leading to a much wider view and a  longer lifetime of the signal. This long life visibility it has been
 observed by recent  brightening of radio and X GRB170817A afterglow, seven month later.

 The solid angle of the thin disk it is still small.
 If we imagine anyway that the gamma disk it is trembling and precessing as the inner jet the average
 solid angle may be spread in a wide belt solid angle (see last figure below), making its detection almost as easy as a spherical one.

Indeed since six months of GRB 170817a radio persistence and its brightening  after its explosion make the event the most outsider of  the GRBs.
However in order to converge the poor statistical probability  (even for a side view jet) into a more sound and larger value, we  suggested that the initial base of the circular jet spiralling electrons does shine (not only) in its  final thin collimated beam, but the jet shined also in a very wide spread gamma synchrotron disk, orthogonal to the same narrow jet. This new morphology (gamma disk and narrow late beamed jet) offers a  wider solid angle view  (and higher probability detection) for nearby distances and a more solid understanding of recent GW-GRB connection, while keeping memory of the past cosmic GRB with a more beamed nature.

 This very lateral $side$ view geometry, better tuned also with the GRB duration,  has been very recently confirmed
 also as an off-axis afterglow light curves and images from 2D hydrodynamic
simulations of double-sided GRB jets in a stratified external medium \cite{Granot et al. (2018)}.

In a sentence, we very probably observed the GW-GRB almost in orthogonal side of the jet.
However most of the present (still popular) Jet model for GRB are just a single shoot event without any additional survival or persistence.
On the contrary our thin persistent GRB jet model \cite{1999A&AS..138..507F},\cite{2001foap.conf..347F},\cite{Fargion:2006we},\cite{Fargion:2016} it is based on an accretion disk
relic of a cannibal collapse (NS-NS or NS-BH) able to feed for long time a jet ,  whose power
may even apparently re-brightening  by a geometry bending and a more collimation to the observer.
Such a persistence indeed it is apparent in GRB170817A, \cite{Bing Li et al.(2018)}.
This lightening model (in axis or just side lateral view) survive also if we consider the $\gamma$ disk shining while
its main jet it is being spinning,  bent and trembling in different orthogonal directions.
Naturally the polarization of the Jet it is the main probe of its synchrotron nature, while the life persistence it is
the test of a permanent (even if slowly decaying) thin precessing jet and its linked orthogonal gamma disk.
Just to view an analogy we show below the Eta Carina nebula image, whose spread inner twin precessing jet shine an apparent hour glass envelope, while the brightest orthogonal radial spread arrows possibly show the twin
gamma disk, whose  pressure eject gas in a radial disk winds.

\subsection{The NS-NS  thermal MeVs neutrino burst}
A GW-GRB correlation might  in next decade discover a signal by a thermal tens MeV neutrinos born in nearest GW-GRB explosion as soon as Hyper-Kamiokande (in Japan and in Korea)
 future Megaton detectors, may reach a threshold of detection at a few Mpc distances.
  The NS-NS neutrino emission may be comparable or just less than a SN one \cite{Fargion:2016}.
  LIGO-VIRGO array may soon increase their sensitivity able to reveal also  a GW by anisotropic SN collapse in few Mpc distances.
The few Mpc size imply a very rare detection of similar NS-NS in next decades, but a well probable one in case of an
anisotropic SN explosion by tens MeV neutrino and by kilohertz GW
possibly recordable by improved LIGO-VIRGO or even LISA antennas.
Indeed the NS-NS emission will be (very probably) ejecting in a spherical symmetry as much as energy as in GW  by thermal neutrinos \cite{Fargion:2016}.
In analogy the same NS-NS collapse neutrino signal, from similar distance as GRB 170817a might rise soon at ten Mpc in a  more large and crowded neutrino array  detector
as an implemented Gigaton one, able to observe neutrinos also at tens MeV energy thresholds.

\section{Conclusions}

The GW astronomy is a great piece in the widest astronomy puzzle. The LIGO-VIRGO discover
has been a great achievement. Its understanding it is not obvious and the GRB morphology it is not a usual GRB blazing lighthouse.

The possibility that future LIGO-VIRGO tens solar mass BH-BH merging are
 catalyzed along dense AGN BH accretion disk might (as in extreme heavy ones \cite{Fargion:2016b}) soon or later
been revealed also by additional Kepler and Lorentz viscous imprint inside the detailed GW tail  structure. For instance
 additional peculiar Shapiro phase delay within the GW structure \cite{FAR.CONV.SHAPIRO.96} or viscous forces during the BH-BH GW emissions.
 In a such a rush and a search for an astronomy, as we mentioned above,
there is room to search for NS-NS or light BH merging at Kilo-Hertz also for a rare GWs conversion ( by galactic or planetary magnetic fields) into kHz radio waves \cite{Fargion,1991}, \cite{Fargion1995}, \cite{Dolgov.et.al.(2017)}; in particular by searching for correlated GW event in radio bangs  captured in the lowest frequency military radio band records, low frequency used for most secret submarine communicate systems.

 The very recent lucky GW-GRB 170817A event, its first kilonova explosion (possibly related to a minor relic NS fragment explosion \cite{Fargion:2016}) and the optical transient allowed anyway to disentangle a very first tag to a sharp GW astronomy. The proposed   gamma disk morphology orthogonal to the thin GRB precessing jet
 might take place and it could enlarged its detection; such a spread shining promises more and more of their discoveries in nearby space (few tens Mpc)  as soon as the GW array antenna will be back at record.
  This very lateral $side$ view geometry, better tuned also with the GRB duration,  has been very recently confirmed
 also as an off-axis side view afterglow light curves and images from 2D hydrodynamic
simulations of double-sided GRB jets in a stratified external medium \cite{Granot et al. (2018)}.
The more collimated gamma thin jet in GRB arise at later advanced relativistic spiral turns more vertical ones: their narrow cones and solid angle shine brighter from far distant edges and  from
larger cosmic volumes.
 This peculiar GRB geometry may also offer a reading key of  known but un-explained relic nebulae, like the well known Eta Carina or Crab one, a jet possibly fed by a persistent accretion disk falling into a NS or  BH.

\section*{Acknowledgements}
The work by MK was supported by Russian Science Foundation
and full fill  in the framework of MEPhI Academic Excellence Project
(contract 02.a03.21.0005, 27.08.2013).

\end{document}